\documentclass{article}

\usepackage{amsmath,amssymb}
\usepackage[dvips]{graphicx}

\makeatletter
\renewcommand{\@cite}[2]{(#2)}

\makeatother

\begin{document}

\begin{center}
{\Large {\bf Albert Einstein's
1916 Review Article on General Relativity}\footnote{To appear in:
{\it Landmark Writings in Western Mathematics, 1640--1940}, Ivor
Grattan-Guiness (ed.), Elsevier 2004.} 

\bigskip
Tilman Sauer\\
\bigskip
{\footnotesize
{ Einstein Papers Project}\\
{ California Institute of Technology 20-7}\\
{ Pasadena, CA 91125, USA}\\
{ Email: tilman@einstein.caltech.edu}} }
\end{center}

\bigskip\noindent
{\bf Abstract}\\[0.3cm]
The first comprehensive overview of the final version of the
general theory of relativity was published by Einstein in 1916
after several expositions of preliminary versions and latest
revisions of the theory in November 1915. A historical account of
this review paper is given, of its prehistory, including a
discussion of Einstein's collaboration with Marcel Grossmann, and
of its immediate reception.
\bigskip


\noindent {\it First publication.} {\it Annalen der Physik, 49},
769--822; also published separately as Leipzig: Barth, 1916.

\bigskip
\noindent {\it Later editions.} Various reprints of the separately
printed version, 5th unaltered reprint in 1929; included in the
3rd and later editions of the anthology Lorentz, H.A., Einstein,
A., Minkowski, H., {\it Das Relativit\"atsprinzip}, Leipzig:
Teubner, 1919$^3$; in the 1960 Readex Microprint edition of the
{\it Published Writings of Albert Einstein} as item~78; in the
anthology K.v.~Meyenn (ed.), {\it Albert Einstein's
Relativit\"atstheorie. Die grundlegenden Arbeiten}, Braunschweig:
Vieweg, 1990; first edition reprinted with annotation as Doc.~30
in Vol.~6 of the {\it Collected Papers of Albert Einstein},
Princeton: Princeton University Press, 1996, pp.~283--339; the
German and English reprints of the {\it Collected Papers} are also
available online at http://www.alberteinstein.info (2003).

\bigskip
\noindent {\it English translations.} 1920 by S.~N.~Bose, in: {\it
The Principle of Relativity. Original Papers by A. Einstein and
H.~Minkowski}, Calcutta: University of Calcutta Press,
pp.~90--163; 1923 by W.~Perrett and G.~B.~Jeffery (without the
first page), in English edition of Lorentz, H.~A.\ et al., {\it
The Principle of Relativity}, London: Methuen 1923 (Dover reprint
1952), pp.~111--164.

\bigskip
\noindent {\it French translations.} 1933 by M.~Solovine, in:
Albert Einstein. {\it Les Fondements de la Th\'eorie de la
Relativit\'e G\'en\'erale. Th\'eorie unitaire de la Gravitation et
de l'Electricit\'e. Sur la Structure Cosmologique de l'Espace.}
Paris: Hermann, pp.~7--71; 1993 by F.~Balibar et al., in {\it
Albert Einstein. Oeuvres choisies}, tome 2, Paris: \'Editions du
Seuil, \'Editions du CNRS, pp.~179--227.

\bigskip
\noindent {\it Russian translation.} 1935, in {\it Albert
Einstein. Sobranie naychnykh trudov}, vol.~1, Moscow: Izdatel'stvo
`Nauka', 1965, pp.~452--504.

\bigskip
\noindent {\it Spanish translation.} 1950 by F.~Alsina Fuertes and
D.~Canals Frau, in Albert Einstein. {\it La relatividad (Memorias
originales)}, Buenos Aires: Emec\'e editores, pp.~115--223.

\bigskip
\noindent {\it Manuscripts.} A manuscript of 46 pages is in the
Schwadron collection at the Hebrew University Jerusalem, available
online at http://www.alberteinstein.info under Call Nr.\ 120-788.

\bigskip

\begin{center}
{\bf 1.~The special theory of relativity}
\end{center}

Some ten years before the first review of the {\it general} theory
of relativity, Einstein published his famous paper {\it On the
Electrodynamics of Moving Bodies} \cite[Einstein
1905]{Einstein1905r}. That paper introduced what later became to
be called the {\it special} theory of relativity. It presented a
conceptual analysis of the notions of space and time, with a
critical reassessment of the meaning of simultaneity at its core.
Length contraction and time dilation in a system that is in
uniform relative motion to an observer with a speed comparable to
that of light are its most salient features.

The 1905 paper was not a very sophisticated paper on the
mathematical side. Its author had obtained a diploma as secondary
school teacher for mathematics and physics at the Polytechnic
Zurich in 1900 \cite[Pais 1982]{Pais1982}, \cite[F\"olsing
1998]{Foelsing1998}. His science education had been excellent with
laboratory work in the most up-to-date facilities and first-rate
mathematics teachers, like Adolf Hurwitz (1859--1919), Carl
Friedrich Geiser (1843--1934), and Hermann Minkowski (1864--1909).
If more recent advances in theoretical physics were somewhat
neglected by his physics teacher Heinrich Friedrich Weber
(1843--1912), the young Einstein made up for it in extensive
autodidactic studies. Fascinated by laboratory experience,
Einstein seems to have skipped more than one of his mathematics
lectures, though, and obtained his knowledge when preparing for
examinations with the help of lecture notes that had been
carefully worked out by his more mathematically inclined friend
Marcel Grossmann (1878--1936).

After initial attempts to start a traditional academic career had
failed, Einstein composed his theory of special relativity in the
evening hours after office work as a technical expert, especially
for electrotechnology, at the Patent office in Bern.
Mathematically, the breakthrough of special relativity came in a
representation using only standard techniques of elementary
calculus. Maxwell's electromagnetic equations were written
component-wise, notwithstanding the fact, that compact vector
notation had already been well developed, if not standardized, in
electrodynamics and hydrodynamics by the end of the nineteenth
century.

The subsequent generalization of the special theory of relativity
to a generally covariant theory of gravitation proceeded in three
major steps \cite[Norton 1984]{Norton1984}, \cite[Stachel
1995]{Stachel1995}, \cite[Renn and Sauer 1999]{RennSauer1999},
\cite[Stachel 2002, sec.~V]{Stachel2002}, \cite[Renn et al.
forthcoming]{Rennetal}. For further references, see the literature
cited in these works, and, on specific aspects, see also volumes 1
\cite[Howard and Stachel 1989]{ES1}, 3 \cite[Eisenstaedt and Kox
1992]{ES3}, 5 \cite[Earman, Janssen and Norton 1995]{ES5}, and 7
\cite[Goenner et al. 1999]{ES7} of the Einstein Studies series.
These steps are:
\begin{itemize}
\item the formulation of the {\it equivalence hypothesis} in 1907,
\item the introduction of the {\it metric tensor} as the crucial
mathematical concept for a generally relativistic theory of
gravitation in 1912,
\item and the discovery of the generally covariant {\it field equations
of gravitation} in 1915.
\end{itemize}

\begin{center}
{\bf 2.~The equivalence hypothesis}
\end{center}

In 1907, Einstein saw himself confronted with the task of
reflecting on the consequences of the relativity principle for the
whole realm of physics. He was asked to write a review article
{\it On the Relativity Principle and the Conclusions Drawn from
It} \cite[Einstein 1907]{Einstein1907j}. The reinterpretation of
the concept of simultaneity in special relativity was hinging on
the finiteness of the speed of light for signal transmission. It
was therefore clear that the Newtonian theory of gravitation posed
an embarrassment. In Newtonian mechanics, the gravitational force
is an action-at-a-distance force and thus contradicts the
fundamental assumption of special relativity that no physical
effects can propagate with a speed superseding a finite value. In
reflection on this difficulty, Einstein took a decisive turn. He
linked the problem of the instantaneous propagation of the
gravitational force in Newtonian physics to the problem of
generalizing the principle of (special) relativity to non-uniform
relative motion. In a reinterpretation of Galileo's law of free
fall, according to which all bodies in a gravitational field
undergo the same acceleration regardless of their weight, Einstein
formulated the so-called equivalence hypothesis. According to this
hypothesis, there is no conceivable experiment that could
distinguish between processes taking place in a static and
homogeneous gravitational field and those that are only viewed
from a frame of reference that is uniformly and rectilinearly
accelerated in a gravitation free space. The value of this
hypothesis was a heuristic one. It enabled Einstein to investigate
the effects of gravitation in a relativistic theory by analyzing
the corresponding processes if interpreted from an accelerated
frame of reference.

Already in 1907, Einstein drew three important consequences from
the equivalence hypothesis. He concluded that the time and hence
also the speed of light must depend on the gravitational
potential. Consequently, the frequency of light emitted from the
sun should be shifted towards the red, and light rays passing
through a gravitational field would be bent. He also concluded
that every energy should have not only inertial but also
gravitational mass.

Incidentally, this is also the time when Einstein began to use the
term `relativity theory' ({\it Relativit\"atstheorie}) in print,
e.g.~\cite[Einstein 1907, p. 439]{Einstein1907}. The term had
first been used in print in the same year by Paul Ehrenfest
(1880--1933), after Max Planck (1858--1947) had earlier introduced
the term {\it Relativtheorie}. A suggestion by Felix Klein
(1849--1925) in 1910, to use the perhaps more appropriate term
`invariant theory' ({\it Invariantentheorie}) was not taken up
\cite[{\it The Collected Papers of Albert Einstein} (CPAE), Vol.
2, p.~254]{CP2}.

While the equivalence hypothesis of 1907 provided a point of
departure for a generalization of the theory of relativity and for
a new field theory of gravitation, Einstein did not present a
solution to the problem of instantaneous propagation of the
gravitational force. While Einstein remained rather silent on the
topic of the relativity principle for some years, these questions
were taken up by others. Hermann Minkowski and Henri Poincar\'e,
e.g., proposed Lorentz-covariant generalizations of Newton's law
of gravitation. More importantly, Minkowski also gave the theory
of relativity a more sophisticated mathematical representation.
Reflecting on the symmetry of the Lorentz transformations,
Minkowski used elements from Cayley's matrix calculus to give the
equations a four-dimensional representation and to interpret the
Lorentz transformations as rotations in a four-dimensional vector
space \cite[Minkowski 1908]{Minkowski1908}. In a report of his
work to the 80th general assembly of physicians and scientists in
Cologne, he illustrated this interpretation by the often-quoted
words:
\begin{quote}
From this hour on, space by itself, and time by itself, shall be
doomed to fade away in the shadows, and only a kind of union of
the two shall preserve an independent reality.
\cite[Minkowski 1909, 105]{Minkowski1909}
\end{quote}
Minkowski's four-dimensional representation was taken up by Arnold
Sommerfeld (1868--1951) who developed a four-dimensional vector
algebra and vector calculus and by Max Laue (1879--1960) who
focused upon the tensorial representation of the
stress-energy-momentum complex.

\begin{center}
{\bf 3.~The metric tensor}
\end{center}

Einstein resumed work on the subject again in 1911. By then he had
been appointed ordinary professor of physics at the German
university in Prague. In a series of papers, he developed a theory
of the static gravitational field, following the heuristics of the
equivalence assumption of static homogeneous gravitational fields
to systems in uniform and rectilinear acceleration \cite[Einstein
1911]{Einstein1911h}, \cite[1912a]{Einstein1912c},
\cite[1912b]{Einstein1912d}. His work was boosted by a competition
with Max Abraham (1875--1922) who had picked up on Einstein's idea
of a variable speed of light and had suggested a dynamic theory of
gravitation. Abraham had proposed a field equation where the
d'Alembertian acting on the speed of light $c$ was proportional to
the scalar mass density. In the course of the debate it quickly
became clear that with variable $c$ Abraham's equation was Lorentz
covariant at best in some ill-defined infinitesimal sense and
could hardly be interpreted consistently. But Abraham had
demonstrated to Einstein the technical power of a four-dimensional
representation, and had prepared him to take the second big step
of introducing the metric tensor.

The second indication of where to go next in the course of
generalizing the relativity principle came from the analysis of
rotating frames of reference. The heuristic assumption of the
equivalence hypothesis implied that also centrifugal and Coriolis
forces should be interpreted as gravitational forces. Looking at
the invariant $c^2dt^2 - (dx^2+dy^2+dz^2)$ in rotating frames of
reference would produce terms of the form $2\omega dx'dt'$ where
the angular velocity $\omega$ would have to be interpreted as a
gravitational potential, just as in the theory of static
gravitation the speed of light $c=c(x,y,z)$ had assumed the role
of a variable gravitational potential. Since moreover the
measuring rods for determining the circumference, but not the
diameter, of a rotating disk are Lorentz contracted, the analysis
of a rotating disk already pointed to a breakdown of Euclidean
geometry.

\begin{center}
{\bf 4.~Einstein's collaboration with Marcel Grossmann}
\end{center}

At some point around this time, Einstein remembered Geiser's
lectures on Gaussian surface theory which he had studied through
his friend's Grossmann's notes. It occurred to him that the
invariant line element of differential geometry might be the key
to finding a proper mathematical representation for his problem.
Fortunately, Einstein had just accepted a call to the Zurich
Polytechnic where Grossmann had become professor of geometry in
1907. Einstein asked Grossmann for help in studying the
mathematical literature, and the two embarked on an intense
collaboration. About this collaboration, he wrote in October 1912:
\begin{quote}
I am now working exclusively on the gravitation problem and
believe that I can overcome all difficulties with the help of a
mathematician friend of mine here. But one thing is certain: never
before in my life have I troubled myself over anything so much,
and I have gained enormous respect for mathematics, whose more
subtle parts I considered until now, in my ignorance, as pure
luxury.
\cite[CPAE, Vol.~5, Doc.~421]{CP5}.
\end{quote}
The question that Einstein put to Grossmann was to identify the
mathematics connected with the invariance of a four-dimensional
infinitesimal line element with metric tensor $g_{\mu\nu}$
\begin{equation}
ds^2 = \sum_{\mu\nu=1}^4g_{\mu\nu}dx^{\mu}dx^{\nu}.
\end{equation}
A research notebook with calculations from that time documents
Einstein's and Grossmann's cooperation \cite[Norton
1984]{Norton1984}, \cite[Renn and Sauer 1999]{RennSauer1999},
\cite[Renn et al. forthcoming]{Rennetal}. It is in this so-called
`Zurich notebook', that we find the first written instance of the
metric tensor for (3+1)-dimensional space-time \cite[Renn and
Sauer 1999, 96]{RennSauer1999}, see also Call No.~3-006, image~39,
on http://www.alberteinstein.info (2003) for a facsimile.
Realizing that the vector calculus for Euclidean space in
curvilinear coordinates is formally equivalent to the calculus of
a general manifold equipped with an invariant infinitesimal line
element, Grossmann saw that the task was to generalize the
four-dimensional vector calculus developed by Minkowski,
Sommerfeld, Laue, and others using methods of an altogether
coordinate independent calculus. Scanning the literature,
Grossmann soon found the necessary mathematical concepts in
\cite[Riemann 1892]{Riemann1892} on $n$-dimensional manifolds, in
\cite[Christoffel 1869]{Christoffel1869} on quadratic differential
forms, and in \cite[Ricci and Levi-Civita 1901]{RicciLeviCivita}
on their so-called absolute differential calculus.

It seems that Einstein and Grossmann quickly saw how to formulate,
in outline, a generally covariant theory with the metric tensor
$g_{\mu\nu}$ representing the gravito-inertial field. In the
following discussion, I will give all formulas in a notation that
is both slightly modernized and made consistent over the various
texts discussed. In particular, I will abbreviate coordinate
derivatives by subscript commas, use the Einstein summation
convention of summing over repeated indices, and denote functional
derivatives by $\delta$ rather than $\partial$. In their joint
publications, Einstein and Grossmann also used Greek letters to
denote contravariant vectors and tensors rather than superscript
indices.

Einstein and Grossmann found generally covariant equations of
motion of a material point of invariant mass $m$ for a given
metric field $g_{\mu\nu}$ in the absence of non-gravitational
forces as
\begin{equation}
\delta\left\{\int Ldt\right\} = \delta\left\{-m\int ds\right\}=0,
\end{equation}
with a particle Lagrangian $L = -mds/dt$. In a generalization to a
continuous distribution of matter characterized by an
energy-momentum tensor for pressureless flow of dust with rest
mass density $\rho_0$,
\begin{equation}
T^{\mu\nu} = \rho_0 \frac{dx^{\mu}}{ds} \frac{dx^{\nu}}{ds},
\end{equation}
the equation of motion turned into
($g=\operatorname{det}(g_{\mu\nu})$)
\begin{equation}
\left(\sqrt{-g}g_{\sigma\mu}T^{\mu\nu}\right)_{,\nu} -
\frac{1}{2}\sqrt{-g}g_{\mu\nu,\sigma}T^{\mu\nu} = 0.
\label{eq_Tdiv}
\end{equation}
The latter equation is an explicit expression for the vanishing of
the covariant divergence of the mixed tensor density
$\sqrt{-g}T^{\nu}_{\sigma}$. It is as such closely related to the
conservation of energy-momentum as can be seen by integrating
$T^{\mu\nu}$ over a closed 3-surface and invoking Gauss's theorem.
In Einstein's interpretation, the first term of (\ref{eq_Tdiv})
gave the conservation law for special relativity for constant
$g_{\mu\nu}$, and the second part consequently represented the
energy-momentum flow due to the gravitational field. This
interpretation led Einstein to believe that the gravitational
force components are given by $g_{\mu\nu,\sigma}$. The task
remained to find a field equation for the metric tensor field,
i.e.\ a tensorial generalization of the Poisson equation.

\begin{center}
{\bf 5.~Coming close to the solution, or so it seems}
\end{center}

From Riemann's and Christoffel's investigations, Grossmann and
Einstein learned that the crucial mathematical concept was the
Riemann curvature tensor $\left\{ik,lm\right\}$ given in terms of
the Christoffel symbols of the second kind (given in the
old-fashioned notation),
\begin{equation}
\left\{\begin{matrix}\mu \; \nu\\ \tau\end{matrix}\right\} =
g^{\tau\lambda}\left(g_{\mu\lambda,\nu} + g_{\nu\lambda,\mu} -
g_{\mu\nu,\lambda}\right),
\end{equation}
as
\begin{equation}
\left\{\iota\kappa,\lambda\mu\right\} = \left\{\begin{matrix}\iota \; \lambda\\
\kappa\end{matrix}\right\}_{,\mu} - \left\{\begin{matrix}\iota \; \mu\\
\kappa\end{matrix}\right\}_{,\lambda} + \left\{\begin{matrix}\iota \; \lambda\\
\rho\end{matrix}\right\} \left\{\begin{matrix}\rho \; \mu\\
\kappa\end{matrix}\right\} - \left\{\begin{matrix}\iota \; \mu\\
\rho\end{matrix}\right\} \left\{\begin{matrix}\rho \; \lambda\\
\kappa\end{matrix}\right\},
\end{equation}
(see Figure 1). Since the right-hand side of the field equation
would be given by the stress-energy tensor of matter, a tensor of
second rank, the left-hand side of the field equation also had to
to be a two-index object. But the obvious candidate, the Ricci
tensor
\begin{equation}
R_{\mu\nu} = \left\{\mu\kappa,\kappa\nu\right\},
\end{equation}
would not produce a field equation that was acceptable to Einstein
and Grossmann at the time. Although a field equation,
\begin{equation}
R_{\mu\nu} + \kappa T_{\mu\nu} = 0, \label{eq_Ricci}
\end{equation}
with some constant $\kappa$ was considered as a candidate, they
dismissed it because they were unable to recover familiar
Newtonian physics in the weak field limit
$g_{\mu\nu}=\eta_{\mu\nu} + h_{\mu\nu}$ with $\eta_{\mu\nu} =
\operatorname{diag(1,1,1,-1)}$, $|h_{\mu\nu}|\ll1$, and
$|h_{\mu\nu,\rho}|\ll 1$.

\begin{figure}
\begin{center}
\includegraphics[scale=.6]{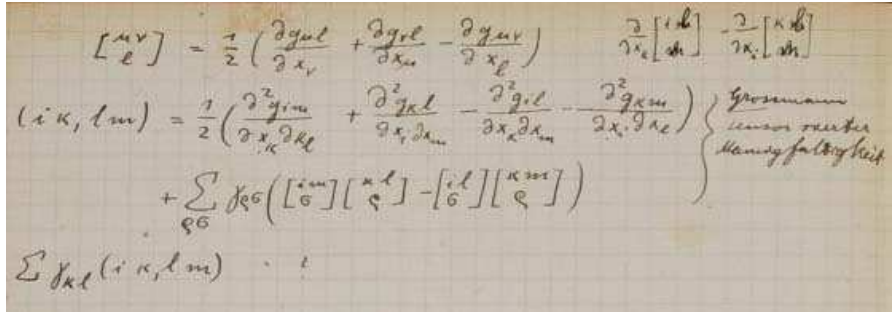}
\caption{Top portion of p.~14L of the `Zurich Notebook' (Einstein
Archives Call.No. 3-006). Next to Grossmann's name Einstein writes
down the Christoffel symbols of the first kind, and the fully
covariant Riemann tensor $(ik,lm)$ which he calls a ``tensor of
fourth manifold'' ({\it Tensor vierter Mannigfaltigkeit}).
Einstein then begins to investigate the Ricci tensor by
contracting with the contravariant metric $\gamma_{kl}$.
\copyright~The Hebrew University of Jerusalem, Albert Einstein
Archives. Reproduced with permission.}
\end{center}
\end{figure}

The dismissal of the candidate (\ref{eq_Ricci}) has been a major
puzzle for historians for a long time. Since in the vacuum case,
$T_{\mu\nu}\equiv 0$, (\ref{eq_Ricci}) is equivalent to the final
field equations of general relativity (see (\ref{eq_Nov4}) below),
Einstein and Grossmann had come by a hair's breadth to arriving at
general relativity already at this point, or so it seems. However,
a closer analysis of the Zurich notebook revealed that Einstein
had to overcome more conceptual difficulties before he was ready
to accept a generally covariant theory \cite[Renn and Sauer
1999]{RennSauer1999}, \cite[Renn et al. forthcoming]{Rennetal}.

\begin{center}
{\bf 6.~The {\it Entwurf} theory}
\end{center}

After giving up the attempt to base a field equation on the
Riemann curvature tensor, Einstein and Grossmann constructed a
field equation that was closer to their heuristic requirements of
energy conservation and recovery of the Poisson equation in the
Newtonian limit. The idea was to take the expression
$\left(g^{\alpha\beta}g^{\mu\nu}_{,\beta}\right)_{,\alpha}$ which
would clearly reduce to the d'Alembertian and Laplacian operators
in the weak field and static limits and substitute it for
$T^{\mu\nu}$ in the second term of (\ref{eq_Tdiv}). If additional
terms of higher order could be identified such that this
expression could be transformed into a total divergence,
energy-momentum conservation in the form of (\ref{eq_Tdiv}) would
automatically be satisfied. The field equations they found read
\begin{align}
\frac{1}{\sqrt{-g}}\left(\sqrt{-g}g^{\alpha\beta}g^{\mu\nu}_{,\beta}\right)_{,\alpha}
-
g^{\alpha\beta}g_{\tau\rho}g^{\mu\tau}_{\alpha}g^{\nu\rho}_{\beta}
&+ \notag\\
+
\frac{1}{2}g^{\alpha\mu}g^{\beta\nu}g_{\tau\rho,\alpha}g^{\tau\rho}_{,\beta}
&-
\frac{1}{4}g^{\mu\nu}g^{\alpha\beta}g_{\tau\rho,\alpha}g^{\tau\rho}_{,\beta}
= -\kappa T^{\mu\nu}.
\end{align}

In early summer 1913, Einstein and Grossmann proceeded to publish
their findings in a little booklet under the title {\it Outline}
[Entwurf] {\it of a Generalized Theory of Relativity and of a
Theory of Gravitation} \cite[Einstein and Grossmann
1913]{EinsteinGrossmann1913}. As the title page indicated, it was
divided into two parts, a physical part for which Einstein signed
responsible, and a mathematical part for which Grossmann signed as
author.

The {\it Entwurf} theory, as it is frequently called in modern
historical literature, was a hybrid theory, if viewed from our
modern understanding of general relativity. It presented a
mathematical apparatus of tensor calculus that allowed to
formulate a theory in a generally covariant manner, and it gave
generally covariant equations of motion. Just as in the final
theory of general relativity, the crucial concept was the metric
tensor which was interpreted as representing a gravito-inertial
field. All these elements were later to be found in the final
version of general relativity. The only thing that was missing
were generally covariant field equations.

The hybrid character of the {\it Entwurf} theory is reflected in a
certain ambivalence that Einstein showed with respect to their
achievement. Initially and also again and again over the following
two years he expressed himself rather pleased with the theory. He
had settled on the {\it Entwurf} equations as acceptable equations
and began to elaborate their consequences. From an unpublished
manuscript we know that together with his friend Michele Besso
(1873--1955) he calculated the advance of the planetary perihelia.
For Mercury, it was well known that the observed perihelion
advance was in discrepancy with the value calculated on the basis
of Newtonian mechanics, and this anomaly was the most prominent
quantitative failure of classical gravitation theory. Not
surprisingly, they found a value for Mercury that was
significantly off the observed value: theirs even came with the
wrong sign \cite[Earman and Janssen 1993]{EarmanJanssen1993}.

Notwithstanding Einstein's acceptance of the {\it Entwurf}
equations, he also indicated that the restricted covariance of
these equations was a `black spot' of the theory. His initial
heuristics clearly did not imply any reason for a restricted
covariance of the theory. In further reflection, Einstein
convinced himself, however, that this restricted covariance was,
in fact, to be expected. He devised an argument to the effect that
indeed no generally covariant field equations were physically
admissable. The argument was first published in an addendum to a
reprint of the {\it Entwurf} in the {\it Zeitschrift f\"ur
Mathematik und Physik}.

He considered a hole in four-dimensional space-time, i.e.\ a
finite region with vanishing stress-energy $T_{\mu\nu}\equiv 0$.
Let $G(x)$ denote a solution $g_{\mu\nu}(x_1,x_2,x_3,x_4)$ of the
field equations, and perform a coordinate transformation within
the hole, i.e.\ consider a coordinate system $x'$ that coincides
smoothly with the original coordinate system $x$ at the boundary
of the hole. In the primed coordinates the transformed field
$G'(x')$ is the solution to the transformed field equations. But
if the field equations are generally covariant, then $G'(x)$ is
also a solution to the original field equations. We hence arrive
at two distinct solutions in the same coordinate system $x$ for
the same distribution of matter $T_{\mu\nu}$. Einstein concluded
that generally covariant field equations cannot uniquely determine
the physical processes in a gravitational field. Consequently, one
had to restrict the admissible coordinate systems to what he began
to call `adapted coordinates'.

Already in their {\it Entwurf}, Einstein and Grossmann had stated
that the most urgent unsolved problem of their theory was the
identification of the covariance group of their field equations.
The solution to this question was made possible by a variational
reformulation of the theory. It was the topic of Einstein's and
Grossmann's second joint publication \cite[Einstein and Grossmann
1914]{EinsteinGrossmann1914b}.

As acknowledged in a footnote, the hint of trying a variational
approach came from Paul Bernays (1888--1977), a student of David
Hilbert (1862--1943) in G\"ottingen. The idea was that a
variational formulation might help to identify the group of
`adapted coordinates' since it would be easier to identify the
invariance group of the scalar action integral than the covariance
group of the explicit tensorial field equations. Einstein and
Grossmann indeed succeeded to cast the {\it Entwurf} theory in a
variational formulation,
\begin{equation}
\delta\left\{\int Ld^4x\right\} = 0,
\end{equation}
with a Lagrangian
\begin{equation}
L =
\sqrt{-g}\left(\frac{1}{4}g^{\alpha\beta}g_{\tau\rho,\alpha}g^{\tau\rho}_{\beta}
-\kappa L^{\rm (mat)}\right), \label{eq_LEntw}
\end{equation}
where the matter part $L^{(\rm mat)}$ was not included explicitly.

Considering variations adapted to the hole consideration, they
were now able to identify the condition for `adapted coordinates'
governing the covariance group of the {\it Entwurf} as
\begin{equation}
B_{\sigma} =
\left(\sqrt{-g}g^{\alpha\beta}g_{\sigma\mu}g^{\mu\nu}_{,\beta}\right)_{,\nu\alpha}
= 0.
\end{equation}
With their second joint paper, the collaboration between Einstein
and Grossmann came to an end. In spring 1914, Einstein moved to
Berlin taking up a position as member of the Prussian Academy of
Sciences in Berlin, a move which relieved him of his teaching load
as professor at the Zurich polytechnic.

\begin{center}
{\bf 7.~The 1914 review article on the {\it Entwurf} theory}
\end{center}

In summer 1914, Einstein felt that the new theory should be
presented in a comprehensive review. He also felt that a
mathematical derivation of the field equations that would
determine them uniquely was still missing.

Both tasks are addressed in a long paper, presented in October
1914 to the Prussian Academy for publication in its {\it
Sitzungsberichte} \cite[Einstein 1914]{Einstein1914o}. It is
entitled {\it The Formal Foundation of the General Theory of
Relativity} and Einstein thus, for the first time, gave the new
theory of relativity the epithet `general' in lieu of the more
cautious `generalized' that he had used for the {\it Entwurf}.

The paper is divided into five sections, and thus anticipates the
structure of the final 1916 review. An introductory section on the
basic ideas of the theory is followed by a section on the theory
of covariants. This section replaced Grossmann's mathematical part
of the joint {\it Entwurf} paper and gives an account of the
elements of tensor calculus employed in the theory. A third
section discusses the theory for a given metric field. It
introduced the stress-energy-momentum tensor and discussed the
conservation laws associated with the vanishing of its divergence,
as well as the equations of motions and the electromagnetic field
equations.

The fourth section gave a new derivation of the {\it Entwurf}
equations. Einstein here tried to give a derivation that
supposedly rendered them unique. He reiterated the hole
consideration and introduced adapted coordinates. The variation is
now done in a generic manner for the gravitational part $H$ of the
Lagrangian $L$. In order to fix the Lagrangian, Einstein assumes
$H$ to be a homogeneous function of second degree in the
coordinate derivatives $g^{\mu\nu}_{,\sigma}$ of the metric, and
picks from the allowed combinations the one that conforms to the
adapted coordinate condition.

In a final, short section Einstein discussed approximations of the
theory, recovered the Newtonian limit and predicted both
gravitational light bending and red shift.

\begin{center}
{\bf 8.~The demise of the {\it Entwurf} and the breakthrough to
general covariance}
\end{center}

Einstein had known that the {\it Entwurf} equations produced the
wrong perihelion advance for Mercury since 1913. A second set-back
that undermined his confidence in the theory came in spring 1915
when Levi-Civita carefully studied Einstein's long Academy paper
and found fault with its derivation of the field equations. After
an intense epistolary exchange in March and April 1915, Einstein
had to admit that his proof of the tensorial character of the left
hand side of the field equations for admissible coordinate
transformations was incomplete \cite[CPAE, Vol.~8, Doc.~80]{CP8}.

In September 1915, Einstein realized that the Minkowski metric in
rotating Cartesian coordinates is not a solution to the {\it
Entwurf} equations. Earlier checks of this condition appear to
have been flawed by trivial algebraic mistakes that conspired to
convince Einstein of the validity of this heuristic requirement
\cite[Janssen 1999]{Janssen1999}.

The final blow came quickly afterwards when Einstein discovered
that the alleged uniqueness of the field equations in his
derivation of the Academy paper did not hold up.

At this point, Einstein began to reconsider alternatives for the
gravitational field equations. He reflected on considerations that
he had done previously in his search for the {\it Entwurf}
equations. A closer analysis of the Zurich notebook indeed
revealed that in the fall of 1915, Einstein reconsidered the same
candidates for field equations as he had done in 1912 \cite[Norton
1984]{Norton1984}, \cite[Renn and Sauer 1999]{RennSauer1999},
\cite[Renn et al. forthcoming]{Rennetal}. The return to general
covariance is documented in four communications to the Prussian
Academy, presented on November 4, 11, 18, and 25, and each
published a week later in the {\it Sitzungsberichte}.

In the first communication, Einstein announced that he had lost
his faith in the {\it Entwurf} equations and wrote
\begin{quote}
In this pursuit I arrived at the demand of general covariance, a
demand from which I parted, though with a heavy heart, three years
ago when I worked together with my friend Grossmann. As a matter
of fact, we were then quite close to that solution of the problem,
which will be given in the following.
\cite[Einstein 1915a, 778]{Einstein1915f}
\end{quote}

Einstein now split the Ricci tensor into two parts,
\begin{equation}
R_{\mu\nu} = \{\mu\kappa,\kappa\nu\} = N_{\mu\nu} + M_{\mu\nu},
\label{eq_Ricci2}
\end{equation}
where
\begin{equation}
N_{\mu\nu} = - \left\{\begin{matrix}\mu \; \nu\\
\kappa\end{matrix}\right\}_{,\kappa} + \left\{\begin{matrix}\mu \; \kappa\\
\rho\end{matrix}\right\}\left\{\begin{matrix}\rho \; \nu\\
\kappa\end{matrix}\right\},
\end{equation}
and
\begin{equation}
M_{\mu\nu} = - \left\{\begin{matrix}\mu \; \kappa\\
\kappa\end{matrix}\right\}_{,\nu} + \left\{\begin{matrix}\mu \; \nu\\
\rho\end{matrix}\right\}\left\{\begin{matrix}\rho \; \kappa\\
\kappa\end{matrix}\right\}.
\end{equation}
Since $\left\{\begin{matrix}\mu \; \kappa\\
\kappa\end{matrix}\right\} = (\ln\sqrt{-g})_{,\mu}$ is a vector
for all transformations that leave $g$ invariant (unimodular
substitutions), $M_{\mu\nu}$ is a covariant derivative of a
vector, and hence all quantities in (\ref{eq_Ricci2}) are tensors
under such substitutions.

The field equations of the first November communication were now
given as
\begin{equation}
N_{\mu\nu} = -\kappa T_{\mu\nu}. \label{eq_Nov1}
\end{equation}
Even though Einstein explicitly reverted to the general covariance
of the Riemann-Christoffel tensor, the field equations of the
first November communication are not generally covariant but only
covariant under unimodular coordinate transformations.

The restricted covariance is immediately obvious also from the
variational formulation that Einstein provided. Looking again at
the geodesic equation,
\begin{equation}
\frac{d^2x^{\tau}}{ds^2} + \left\{\begin{matrix}\mu \; \nu\\
\tau\end{matrix}\right\} \frac{dx^{\mu}}{ds}\frac{dx^{\nu}}{ds} =
0, \label{eq_geodesic}
\end{equation}
as the equation of motion for a point particle in a given
gravitational field, Einstein now conceived of the negative
Christoffelsymbols $\Gamma^{\sigma}_{\mu\nu} = - \left\{\begin{matrix}\mu \; \nu\\
\sigma\end{matrix}\right\}$ as the components of the gravitational
force rather than the simple coordinate derivatives of the metric
$g_{\mu\nu,\sigma}$. These quantities now entered into the
gravitational part of the Lagrangian as
\begin{equation}
L =
g^{\sigma\tau}\Gamma^{\alpha}_{\sigma\beta}\Gamma^{\beta}_{\tau\alpha}
- \kappa L^{(\rm mat)}. \label{eq_Nov1Lagr}
\end{equation}
(cp.~(\ref{eq_LEntw}). He observed that weak fields now allow to
go to the Newtonian limit, and that the transition to rotating
frames of reference is admissible since the corresponding
coordinate transformations have unit determinant.

Not only was the covariance of the theory restricted to unimodular
transformations, Einstein also showed that energy-momentum
conservation demanded that a coordinate restriction,
\begin{equation}
\left(g^{\alpha\beta}\left[\ln\sqrt{-g}\right]_{,\beta}\right)_{,\alpha}
= -\kappa T, \label{eq_cond1}
\end{equation}
had to be satisfied. Since, in general, the trace of the
energy-momentum tensor $T=g^{\mu\nu}T_{\mu\nu}$ does not vanish,
(\ref{eq_cond1}) implies that coordinates cannot be chosen
arbitrarily. In particular, (\ref{eq_cond1}) implies that one
cannot set $\sqrt{-g}\equiv 1$.

At this point, it needs to be mentioned that Einstein's return to
general covariance in November 1915 was done in a hasty
competition with Hilbert \cite[Sauer 1999]{Sauer1999}. Einstein
had given a series of lectures on the {\it Entwurf} theory in
G\"ottingen earlier in the summer, and Hilbert had then closely
studied Einstein's theory over the fall. Apparently, Hilbert had
found fault with Einstein's derivation of the field equations,
too, and Einstein had heard about Hilbert's criticism through
Sommerfeld \cite[CPAE, Vol.~8, Doc.~136]{CP8}. When he received
proofs of his first November communication, he forwarded them to
G\"ottingen, and it seems that Hilbert responded immediately with
a report about his own progress. Hilbert, at the time, believed in
an electromagnetic world-view and had been working on combining
Einstein's gravitational theory with a generalized version of
Maxwellian electrodynamics suggested by Gustav Mie (1868--1957).
Mie had proposed a theory of matter where non-linear, but
Lorentz-covariant generalizations of Maxwell's equations should
allow for particle-like solutions in the microscopic realm. It
seems likely that Hilbert had informed Einstein about the basic
characteristics of his approach which aimed at a unification of
Einstein's and Mie's theories.

The second of Einstein's four famous November communications, in
any case, discussed the possibility of a purely electromagnetic
origin of matter \cite[Einstein 1915b]{Einstein1915g}. Since in
classical electromagnetism, the stress-energy-momentum tensor
$T^{\mu\nu}$ is given in terms of the electromagnetic field tensor
$F_{\mu\nu}$ as
\begin{equation}
T^{\mu\nu} = \frac{1}{4\pi}\left(F^{\mu\alpha}F^{\nu}_{\alpha}
-\frac{1}{4}g^{\mu\nu}F^{\alpha\beta}F_{\alpha\beta}\right),
\end{equation}
it is readily seen that its trace $T$ vanishes identically.
Einstein now entertained the possibility that on a microscopic
level all matter might be of electromagnetic origin. In this case,
the right hand side of the coordinate condition (\ref{eq_cond1})
would vanish and hence coordinates with constant $g$ would be
admissible. In this case, Einstein argued, one could take the
fully covariant equations
\begin{equation}
R_{\mu\nu} = -\kappa T_{\mu\nu}, \label{eq_Nov2}
\end{equation}
which he had already considered earlier, see (\ref{eq_Ricci}), and
reduce them to the field equations (\ref{eq_Nov1}) by choosing
coordinates for which $g\equiv 1$.

The field equations (\ref{eq_Nov2}) still differ from the final
field equations but for the vacuum case, $T_{\mu\nu}=0$, they are
already equivalent. Einstein therefore was able to compute on the
basis of (\ref{eq_Nov2}) the correct unaccounted perihelion
advance by looking at the field of a point mass in second
approximation. The calculation produced the correct value of
$43''$ per century without any arbitrary or ad hoc assumptions. In
the computation Einstein could take advantage of his having
calculated the advance before for the {\it Entwurf} theory. The
new field equations, in fact, only involved a modification of his
earlier calculations \cite[Earman and Janssen
1993]{EarmanJanssen1993}. Einstein published these results in his
third November communication \cite[Einstein 1915c]{Einstein1915i}.

With the success of the perihelion calculation, the return to
general covariance was definite. The final step \cite[Einstein
1915d]{Einstein1915i} was to add a trace term to the matter tensor
to obtain field equations of the form
\begin{equation}
R_{\mu\nu} = -\kappa
\left(T_{\mu\nu}-\frac{1}{2}g_{\mu\nu}T\right). \label{eq_Nov4}
\end{equation}
With the trace term added, the postulate of energy-momentum
conservation no longer produced a coordinate restriction since it
was now automatically satisfied by ({\ref{eq_Nov4}).

Equations (\ref{eq_Nov4}) are the final field equations of the
generally relativistic theory of gravitation, as we know them
today. They are frequently referred to as the `Einstein equations'
of general relativity.

With the exception of the first November communication, where he
had given the Lagrangian (\ref{eq_Nov1Lagr}) for the field
equations (\ref{eq_Nov1}), Einstein had not discussed the
subsequent field equations in a variational approach. The closure
of providing a variational formulation was contributed by Hilbert
in his own approach to a generally covariant theory of gravitation
and electromagnetism. Since he was being kept informed by Einstein
about the latter's progress, he rushed ahead and presented an
account of his own version to the G\"ottingen Academy for
publication in its {\it Nachrichten} on November 20. Page proofs
of Hilbert's original paper show that the version submitted for
publication on November 20 still differed from the version that
was eventually published. But it did already suggest to base the
theory on a variational principle and emphasized that the
Lagrangian must be a scalar function for general coordinate
transformations.

In the printed version of Hilbert's paper, the Riemann curvature
scalar $R$ is taken to be the gravitational part of the Lagrangian
and it is stated, albeit not derived by explicit calculation, that
a variation of the action
\begin{equation}
\mathcal{A} = \int\sqrt{-g}\left(R-\kappa L^{(\rm
mat)}\right)d^4\tau,
\end{equation}
with respect to the metric tensor components $g^{\mu\nu}$ would
produce the gravitational field equations
\begin{equation}
R_{\mu\nu} - \frac{1}{2}g_{\mu\nu}R = -\kappa
\frac{1}{\sqrt{-g}}\frac{\delta L^{\rm (mat)}}{\delta g^{\mu\nu}},
\label{eq_Nov4Hilb}
\end{equation}
which is an equivalent version of Einstein's field equation
(\ref{eq_Nov4}). (\ref{eq_Nov4Hilb}) may be transformed to
(\ref{eq_Nov4}) by looking at the trace of (\ref{eq_Nov4}) and
substituting $R=-\kappa T$ into (\ref{eq_Nov4Hilb}). The
equivalence then follows from the non-trivial identification of
\begin{equation}
\frac{1}{\sqrt{-g}}\frac{\delta L^{\rm (mat)}}{\delta g^{\mu\nu}}
= T_{\mu\nu}.
\end{equation}
In the latter step, Hilbert and Einstein differed considerably
since Hilbert axiomatically took $L^{(\rm mat)}$ to be a function
exclusively of the electromagnetic potential $A_{\mu}$, the
electromagnetic field $F_{\mu\nu} = A_{\mu,\nu}-A_{\nu,\mu}$, and
the metric tensor components $g^{\mu\nu}$,
\begin{equation}
L^{\rm (mat)} = L^{\rm (mat)}(A_{\mu}, F_{\mu\nu}, g_{\mu\nu}),
\end{equation}
in accordance with his electromagnetic world view. Einstein,
however, had entertained the hypothesis of an electromagnetic
origin of matter only for a few days. With his fourth November
communication at the latest, Einstein had given up that hypothesis
again and was allowing for an unspecified $T_{\mu\nu}$ in his
final version of the theory.

\begin{center}
{\bf 9.~The 1916 review paper}
\end{center}

Ever since Levi-Civita had found a gap in Einstein's covariance
proof of the {\it Entwurf} equations, Einstein had meant to update
or rewrite his 1914 Academy article on the general theory of
relativity. With the return to general covariance, the success of
explaining the perihelion advance of Mercury, and the new field
equations (\ref{eq_Nov4}) of the fourth November communication,
Einstein decided to write an altogether new account of the general
theory of relativity.

The new review was received by the {\it Annalen der Physik} on 20
March 1916, some four months after the last November paper. Its
structure is not much different from the earlier 1914 Academy
article. It is again divided into five sections:
\begin{enumerate}

\item[A.] Fundamental Considerations on the Postulate of
Relativity,

\item[B.] Mathematical Aids to the Formulation of Generally
Covariant Equations,

\item[C.] Theory of the Gravitational Field,

\item[D.] Material Phenomena,

\item[E.] [Newtonian Limit and Observable Consequences].

\end{enumerate}
In an introductory paragraph Einstein called the theory to be
expounded in the review `conceivably the farthest-reaching
generalization' of the special theory of relativity. While the
latter is assumed to be known to the reader, he sets out to
develop especially all the necessary mathematical tools
\begin{quote}
---and I tried to do it in as
simple and transparent a manner as possible, so that a special
study of the mathematical literature is not required for the
understanding of the present paper.
\cite[Einstein 1916, 769]{Einstein1916e}
\end{quote}
Nevertheless, in this first paragraph Einstein did mention
Minkowski's formal equivalence of the spatial and time
coordinates, the investigations on non-Euclidean manifolds by
Gauss, Riemann, and Christoffel, and the absolute differential
calculus of Ricci and Levi-Civita. Echoing a theme of Felix
Klein's but also of later commentators, he wrote that especially
the absolute differential calculus had provided mathematical means
which simply had to be taken up---as if he had not struggled hard
for years to apply them in a physically meaningful way. He also
acknowledged Grossmann's help again in studying the mathematical
literature and in searching for the gravitational field equations.

The first section then introduces the postulate of general
covariance, arguing to a large extent from  purely epistemological
considerations. Einstein denounces the existence of an absolute
space by considering two massive bodies far away both from other
masses and from each other and in relative rotation along their
line of connection. If one body were observed to be of spherical
shape and the other to be an ellipsoid, Newtonian mechanics would
have to attribute the cause for the different shapes in a rotation
relative to absolute space. But this is unsatisfactory because a
causal agent is introduced which itself can never be an object of
causal effect nor of observation. Hence, one is forced to
attribute the cause for this change of shape to the distant masses
of the fixed stars, an argument that follows Mach's critique of
classical mechanics.

The second argument is the equivalence hypothesis based on
Galileo's empirical law of free fall. Next, Einstein discusses the
rotating disk to argue for the fact that in general relativity
coordinates no longer have an immediate metric meaning. A fourth
argument in this section was new and replaced the earlier hole
consideration. The hole argument had supposedly proven that no
generally covariant field equations could be given a physical
meaning in accordance with our notions of causality and the demand
that the field equations are determined uniquely by the
energy-matter distribution. Einstein did not explicitly retract
the argument but gave a new consideration, known as the point
coincidence argument. He argued that what we observe in physical
experiments are always only spatio-temporal coincidences. If all
physical processes would consist in the motion of material points,
we could only observe those events where two or more of their
worldlines coincide. Then the coordinates of the four-dimensional
space-time manifold are merely labels for those coincidences, and
no coordinate system must be preferred over any other. The
implicit objection to the hole argument that invalidates its
conclusion is that the different metric fields $G(x)$ and $G'(x)$
obtained by dragging the metric tensor over the hole, do not, in
fact, represent different physical situations since they agree on
all point coincidences.

In the second, mathematical section, Einstein summarily develops
the elements of tensor algebra and tensor calculus. He introduces
contravariant and covariant vectors and general tensors which are
defined by the transformation laws of their components. He
introduces the algebraic operations of external multiplication and
contraction, and of raising and lowering of indices. Among the
properties of the metric tensors, he discusses the invariance of
the volume element $\sqrt{-g}d^4x$. He repeats the derivation of
the geodesic equation, introduces Christoffel's symbols and
discusses covariant differentiation by considering invariance
along the geodesic line. He mentions the fact that the covariant
derivative of the metric vanishes and derives a number of explicit
formulas for the differentiation of contravariant, covariant and
mixed tensors. The last paragraph introduces the
Riemann-Christoffel curvature tensor and discusses its splitting
into two parts, as in (\ref{eq_Ricci2}). Perhaps the most
noteworthy point of the section, compared to earlier expositions
of the mathematical foundations of general relativity, is what
came to be called the `Einstein summation convention'. It is in
this section that Einstein for the first time in print introduced
the convention that in any tensor expression a summation over two
repeated indices is implied with out writing down the summation
sign.

The third section derives the gravitational field equations. They
are given here as
\begin{align}
\Gamma^{\alpha}_{\mu\nu,\alpha} +
\Gamma^{\alpha}_{\mu\beta}\Gamma^{\beta}_{\nu\alpha} &=
-\kappa\left(T_{\mu\nu}-\frac{1}{2}g_{\mu\nu}T\right),
\label{eq_fieldeq16}\\
\sqrt{-g}&=1. \label{eq_gcond}
\end{align}
Somewhat surprisingly, from a modern point of view, Einstein did
not give the field equations in a generally covariant form.
Instead he fixed the coordinates by condition (\ref{eq_gcond}) in
all equations that he gave in the section. He emphasized, though,
that this is a mere specification of the coordinates introduced
for convenience. The introduction of the field equations, in fact,
proceeded by arguing that the vanishing of the Ricci tensor
$R_{\mu\nu}$ is the unique equation that determines the metric
field in the absence of masses if we demand that the expression
depends only on $g_{\mu\nu}$ and its first and second derivatives
and depends on the latter only linearly. The possibility of adding
a term proportional to $g_{im}R$, equivalent in the vacuum case,
(but not of adding a cosmological term proportional to $g_{im}$)
is mentioned in a footnote.

The Lagrangian for the variational form of the field equations in
vacuum is given as
\begin{equation}
L =
g^{\mu\nu}\Gamma^{\alpha}_{\mu\beta}\Gamma^{\beta}_{\nu\alpha},
\end{equation}
together with the explicit stipulation of condition
(\ref{eq_gcond}). The introduction of the matter term proceeds by
defining the stress-energy complex of the gravitational field as
\begin{equation}
\kappa t^{\alpha}_{\sigma} = \frac{1}{2}\delta^{\alpha}_{\sigma}
g^{\mu\nu}\Gamma^{\alpha}_{\mu\beta}\Gamma^{\beta}_{\nu\alpha} -
g^{\mu\nu}\Gamma^{\alpha}_{\mu\beta}\Gamma^{\beta}_{\nu\sigma},
\end{equation}
an expression which is not a tensor under general coordinate
transformation in accordance with the fact that the field energy
associated with the gravito-inertial field is not a localizable
quantity. Using $t^{\alpha}_{\sigma}$, Einstein rewrote the field
equation (\ref{eq_fieldeq16}) as
\begin{equation}
\left(g^{\sigma\beta}\Gamma^{\alpha}_{\mu\beta}\right)_{,\alpha} =
-\kappa\left(t^{\sigma}_{\mu} -
\frac{1}{2}\delta^{\sigma}_{\mu}t\right),
\end{equation}
and demanded that the non-gravitational energy-momentum tensor
$T^{\sigma}_{\mu}$ enters in the equation on the same footing as
$t^{\sigma}_{\mu}$. The latter requirement is equivalent to
demanding that a divergence equation,
\begin{equation}
\left(t^{\sigma}_{\mu} + T^{\sigma}_{\mu}\right)_{,\sigma} = 0,
\end{equation}
holds for the total energy of the system.

While the derivation of the field equations differs considerably
from earlier accounts, the fourth and fifth sections take up
material from earlier expositions. In these sections, Einstein
discussed Euler's hydrodynamic equation with an energy-momentum
tensor
\begin{equation}
T^{\alpha\beta} = -g^{\alpha\beta}p + \rho
u^{\alpha}u^{\beta}, u^{\alpha} = dx^{\alpha}/ds,
\end{equation}
for non-dissipative, adiabatic liquids, characterized by the two
scalars of pressure $p$ and density $\rho$. Electrodynamics is
governed by Maxwell's equations in generally covariant form, and,
in the last section, Einstein discussed the Newtonian
approximation of weak fields, Minkowski flat boundary conditions,
and slow motion of the particles. In the consideration of the
Newtonian limit the constant $\kappa$ may be related to the
gravitational constant $G$ by comparison with Poisson's equation
as $\kappa = 8\pi G/c^2$. Einstein explains a subtlety of the
Newtonian limit that had played a role in his earlier dismissal of
generally covariant equations. In first Newtonian approximation
only the $g_{44}$ components enter into the equations of motion,
even though the postulate of $\sqrt{-g}=1$ demands that the other
diagonal components are non-trivial of the same order. The
first-order diagonal components, however, do enter into the
geodesic equation for a light ray passing in a centrally symmetric
gravitational field. For this reason, the predicted expression for
the light bending of a light ray grazing the edge of the sun, came
out with a factor of 2, compared to earlier considerations that
were based on the equivalence hypothesis alone. The slowing of
clocks in a gravitational field and the gravitational red shift of
spectral lines is discussed explicitly but the calculation of the
perihelion shift for Mercury obtained in second approximation of a
spherically symmetric field is only mentioned with reference to
the pertinent November communication.

\begin{center}
{\bf 10.~Early reception of the final version of general
relativity}
\end{center}

The first exact solution to the field equations---and to date the
most important one---was found almost simultaneously with
Einstein's 1916 review by the astronomer Karl Schwarzschild
(1873--1916). He computed the static, spherically symmetric field
outside a spherically symmetric mass distribution of total mass
$m$. His solution allowed to compute the light bending of light
rays and the planetary perihelion motion without approximation.
The solution is regular everywhere except at the origin but at a
radius $r_S=2Gm/c^2$, now called the Schwarzschild radius, the
time coordinate changes its sign relative to the spatial
coordinates. This coordinate singularity is responsible for what
came to be known as the black hole horizon and its interpretation
presented a major difficulty for many years.

While more exact solutions were found over the following years,
approximation schemes played an equally important role for an
interpretation of the theory. An approximate solution was
discussed by Einstein in the summer of 1916 in a first paper on
gravitational waves. The existence of gravitational waves was
expected in a field theory of gravitation by analogy to the
electromagnetic case. Einstein's first paper on this topic was
marred by a mistake which made him conclude that waves should
exist that do not transport energy. The error was corrected in a
second paper of 1918. Until now, the topic of gravitational waves
is an active field of research and their existence has been shown
indirectly only in 1974 through the energy loss of binary pulsars
(Nobel prize 1993). Experimental efforts to observe gravitational
waves directly are still underway.

The question of energy transport in gravitational waves is
connected to the question of identifying an expression for the
gravitational field energy and a corresponding conservation law.
The question was debated in the years 1916--1919 by a number of
mathematicians, most importantly by Felix Klein. The final
solution came with Noether's theorems  on the connection of
conservation laws and symmetries of the variational formulation.
These theorems were anticipated for a special case in Hilbert's
1915 paper and published in its general form in 1918 by Emmy
Noether (1882--1935).

Einstein tried to encourage experimental efforts aimed at testing
the two main predictions of the theory. A confirmation of the
gravitational red shift was difficult to determine due to the many
competing effects that result in a shifting or broadening of solar
or stellar spectral lines. An unequivocal confirmation of the
gravitational red shift only came in 1960 in a controlled
terrestrial experiment making use of the M\"ossbauer effect.

But the results of a British expedition led by Arthur Eddington
(1882--1944) to test the predicted gravitational light bending
during a solar eclipse on 29 May 1919 in Sobral, Brazil, and on
the island of Principe in the gulf of Guinea, were established and
reached Europe later in the fall of that year. The results
confirmed Einstein's prediction, and within weeks, Einstein turned
into a world celebrity and the theory of relativity into a
household term.

A popular, non-technical account of both the special and general
theories of relativity that Einstein had written in 1917
\cite[Einstein 1917]{Einstein1917a} became a best-seller. A fourth
edition in 1919 was reprinted in a fifth through tenth edition in
1920 and saw a fourteenth edition in 1922. It was also translated
into many languages. The increased interest in Einstein's theory
is also witnessed by an uncountable number of more or less popular
accounts and other books and articles dealing with relativity. A
bibliography of relativity from 1924 lists close to 4000 entries
\cite[Lecat 1924]{Lecat1924}.

The consequences of both special and general relativity began to
be discussed in many circles. Early interpretations of general
relativity from a philosophical point of view had been published
by Moritz Schlick (1882--1936) and Hans Reichenbach (1891--1953).
In the early 1920's philosophical interpretations of relativity
came to abound, the analysis in \cite[Hentschel
1990]{Hentschel1990} carries a bibliography of over 3000 items.
The public interest in Einstein's new theory was not always
untainted by political partisanry. Antisemitic attacks against
Einstein focussed not only on Einstein's person or on his
political and pacifist stance but targeted his theory as well. As
early as 1920, antisemitically motivated objections against the
theories of relativity were expressed in a public meeting at the
Berlin philharmonic in summer 1920 and again at the first post-war
meeting of the Society of German Scientists and Physicians in Bad
Nauheim in September 1920. On the other hand, Einstein began to be
recognized worldwide as a leading physicist. He received
international invitations and honors, and began to travel
extensively giving talks about his theory at a time when post-war
German science was still boycotted by many scholars and scientific
institutions.

\begin{center}
{\bf 11.~Going on and beyond general relativity}
\end{center}

For Einstein, the victory of the breakthrough to general
covariance in November 1915 was not to be regarded as establishing
a final theory that would not be subject to further revisions.
Already in 1917, he modified the gravitational field equations by
adding a term proportional to $\lambda g_{\mu\nu}$ to
(\ref{eq_Nov4}). The modification was motivated in the context of
a cosmological consideration. Einstein wanted to avoid the
stipulation of boundary conditions at infinity in order not to
have to account for inertial effects that might not have been
caused by masses, in accordance with what he called Mach's
principle. He suggested to consider the cosmological model of a
spatially closed and static universe but had to modify the field
equations by introducing the cosmological constant $\lambda$ in
order to allow for the possibility of such a solution. An
alternate vacuum solution to the modified field equations advanced
by Willem de Sitter (1872--1934) soon showed, however, that the
new field equations did not automatically satisfy Mach's principle
as had been Einstein's hope.

In 1919, Einstein entertained the possibility of a gravitational
field equation where the trace term in (\ref{eq_Nov4}) would be
added with a factor of $1/4$ instead of $1/2$. The modification
was motivated by considerations concerning the constitution of
matter and implies that it is no longer the covariant divergence
of $T_{\mu\nu}$ that is automatically vanishing but rather its
trace. Other modifications of the field equations or
generalizations of the underlying Riemannian geometry were
investigated by Einstein and others in the following decades in
attempts to find a geometrized unification of the gravitational
and electromagnetic fields.

In fact, a geometric interpretation of the general theory of
relativity, if considered at all, originally pertained only to the
geodesic equation. Until 1916, the Riemann and Ricci tensors were
only interpreted as algebraic invariants. A geometric
interpretation in terms of parallel transport of tangent vectors
was elaborated in the following years mainly through the work of
Tullio Levi-Civita and Hermann Weyl (1885--1955).

In the course of elaborating the geometric meaning of general
relativity, it was Hermann Weyl, who took the first steps to go
beyond a purely (semi-)Riemannian framework for general relativity
and, at the same time, first proposed a truly geometrized
unification of the gravitational and electromagnetic fields. First
published in 1918, it was later incorporated into the third
edition of his widely read exposition of general relativity
\cite[Weyl 1918]{Weyl 1918}, \cite[Scholz 2001]{Scholz2001}. In
accordance with more general philosophical concerns about the
foundations of mathematics, Weyl's point of departure was the
observation that in Riemannian geometry, no integrable, or
path-independent comparison of vector directions at different
points of the manifold is possible, whereas the length of a vector
remains unaffected during parallel transport. In order to realize
a true `infinitesimal geometry' ({\it Nahegeometrie}), Weyl in
1918 introduced an additional geometric structure, a length
connection, i.e.\ a linear differential form
$d\varphi=\varphi_idx^i$ that governed the transport of vector
lengths $l$ by the definition $\delta l \equiv (\partial
l/\partial x^i)dx^i + l\varphi_i dx^i\equiv 0$. At the same time,
the Riemannian metric $g_{\mu\nu}$ had to be replaced by the class
of conformally equivalent metrics $[g]$ where two representatives
of a class are connected through $\tilde{g}_{\mu\nu} = \lambda
g_{\mu\nu}$ with a scalar function $\lambda$. For consistency, the
length connection $\varphi$ has to be transformed, too, as
$\tilde{\varphi}_idx^i = \varphi_idx^i - d\log \lambda$. For these
transformations, Weyl introduced the term `gauge transformations.'

The (semi-)Riemannian manifold with metric tensor field
$g_{\mu\nu}$ was hence generalized to a manifold with conformally
equivalent classes $[g]$, $[\varphi]$ of (semi-)Riemannian metrics
and length connections. The geometric meaning of this
generalization was realized by investigating the affine
connection, governing the parallel transport of vectors. It
turned out that the curvature associated with the length
connection, i.e.\ the exterior derivative of $f = d\varphi$, in
coordinates, $f_{ij}=\varphi_{i,j}-\varphi_{j,i}$, could be
interpreted as the representation of the electromagnetic field
tensor \cite[Scholz 2001, esp. pp.~63--69]{Scholz2001}.

Einstein's reaction to Weyl's theory was highly ambivalent.
Fascinated by the mathematical analysis, he quickly pointed out
that the theory was inacceptable from a physics point of view
since it implied, e.g., that the wavelength of light emitted by
radiating atoms should depend on the prehistory of the atom,
contrary to experience. Despite this argument, Weyl's theory
proved extremely influential as the first (more or less)
successful attempt to achieve a geometric unification of the
gravitational and electromagnetic fields. During the twenties,
many attempts were tried to achieve a unification of gravitation
and electromagnetism by generalizing Riemann geometry. These
investigations both stimulated and profited from parallel
developments in differential geometry.

With the advent of quantum mechanics in 1926, the discovery of the
weak and strong interactions and the proliferation of elementary
particles in nuclear and subnuclear physics, the parameters for a
unification program changed drastically. Many aspects of the
original unified field theory program have consequently fallen
into oblivion but the history of modern differential geometry can
hardly be understood without taking into account this context of
searching for generalizations of Riemannian geometry.

In essence, Einstein's general theory of relativity of 1916
remains today's accepted theory of the gravitational field, and
notwithstanding the expectation that a generally relativistic
theory of gravitation should also be quantized---an unsolved
problem until today---, classical general relativity, in the sense
of an exploration of the solutions and implicit consequences of
its gravitational field equations, has been an active field of
research ever since.


\begin{center}
{\bf Bibliography}
\end{center}

\noindent CPAE.\ {\it The Collected Papers of Albert Einstein}.
Princeton: Princeton University Press, 1987ff.

\medskip
\noindent Christoffel, E.B.~1869.\ `Ueber die Transformation der
homogenen Differential\-ausdr\"ucke zweiten Grades'. {\it Journal
f\"ur die reine und angewandte Mathematik, 70}, 46--70.

\medskip
\noindent Earman, J. and Janssen, M.~1993. `Einstein's Explanation
of the Motion of Mercury's Perihelion'. in: \cite[Earman, Janssen
and Norton 1995, 129--172]{ES5}.

\medskip
\noindent Earman, J., Janssen, M. and Norton J. (eds.) 1993. {\it
The Attraction of Gravitation: New Studies in the History of
General Relativity.} Boston, Basel, Berlin: Birkh\"auser [Einstein
Studies 5].

\medskip
\noindent Einstein, A.~1905. `Zur Elektrodynamik bewegter
K\"orper'. {\it Annalen der Physik, 17}, 891--921 [CPAE~2,
275--310.].

\medskip
\noindent --- 1907. `\"Uber das Relativit\"atsprinzip und die aus
demselben gezogenen Folgerungen'. {\it Jahrbuch der
Radioaktivit\"at und Elektronik, 4}, 411--462 [CPAE~2, 432--488.].

\medskip
\noindent --- 1911. `\"Uber den Einflu{\ss} der Schwerkraft auf
die Ausbreitung des Lichtes'. {\it Annalen der Physik, 35},
898--908 [CPAE~3, 485--497.].

\medskip
\noindent --- 1912a. `Lichtgeschwindigkeit und Statik des
Gravitationsfeldes'. {\it ibid., 38}, 355--369 [CPAE~4,
129--145.].

\medskip
\noindent --- 1912b. `Zur Theorie des statischen
Gravitationsfeldes'. {\it ibid., 38}, 443--458 [CPAE~4, 146--164].

\medskip
\noindent --- 1914. `Die formale Grundlage der allgemeinen
Relativit\"atstheorie'. {\it K\"oniglich Preu{\ss}ische Akademie
der Wissenschaften} (Berlin). {\it Sitzungsberichte}, 1030--1085
[CPAE~6, 72--130.].

\medskip
\noindent --- 1915a. `Zur allgemeinen Relativit\"atstheorie'. {\it
ibid.}, 778--786 [CPAE~6, 214--224].

\medskip
\noindent --- 1915b. `Zur allgemeinen Relativit\"atstheorie
(Nachtrag)'. {\it ibid.}, 799--801 [CPAE~6, 225--229.].

\medskip
\noindent --- 1915c. `Erkl\"arung der Perihelbewegung des Merkur
aus der allgemeinen Relativit\"atstheorie'. {\it ibid.}, 831--839
[CPAE~6, 233--243.].

\medskip
\noindent --- 1915d. `Die Feldgleichungen der Gravitation'. {\it
ibid.}, 844--847 [CPAE~6, 244--249.].

\medskip
\noindent --- 1916. `Die Grundlage der allgemeinen
Relativit\"atstheorie'. {\it Annalen der Physik, 49}, 769--822
(also published separately as Leipzig: Teubner) [CPAE~6,
283--339.].

\medskip
\noindent --- 1917. {\it \"Uber die spezielle und die allgemeine
Relativit\"atstheorie.} Braunschweig: Vieweg [CPAE~6, 420--539.].



\medskip
\noindent Einstein, A. and Grossmann, M. 1913. {\it Entwurf einer
verallgemeinerten Relativit\"atstheorie und einer Theorie der
Gravitation.} Leipzig: Teubner [reprinted with an addendum in {\it
Zeitschrift f\"ur Mathematik und Physik, 62} 225--244; CPAE~4,
302--343, 579--582.].

\medskip
\noindent --- 1914 `Kovarianzeigenschaften der Feldgleichungen der
auf die verallgemeinerte Relativit\"atstheorie gegr\"undeten
Gravitationstheorie'. {\it Zeitschrift f\"ur Mathematik und
Physik, 63}, 215--225 [CPAE~6, 6--18.].

\medskip
\noindent Eisenstaedt, J. and Kox, A.J. (eds.) 1992. {\it Studies
in the History of General Relativity}, Boston, Basel, Berlin:
Birkh\"auser [Einstein Studies 3].

\medskip
\noindent F\"olsing, A. 1998. {\it Albert Einstein. A Biography.}
Pinguin paperback.

\medskip
\noindent Goenner, H., Renn, J., Ritter, J., and Sauer, T. (eds.)
1999. {\it The Expanding Worlds of General Relativity}, Boston,
Basel, Berlin: Birkh\"auser [Einstein Studies 7].

\medskip
\noindent Hentschel, K. 1990. {\it Interpretationen und
Fehlinterpretationen der speziellen und der allgemeinen
Relativit\"atstheorie durch Zeitgenossen Albert Einsteins.} Basel,
Boston, Berlin: Birkh\"auser.

\medskip
\noindent Hilbert, D. 1915 `Die Grundlagen der Physik. (Erste
Mitteilung.)' {\it K\"onigliche Gesellschaft der Wissenschaften zu
G\"ottingen. Mathematisch-physikalische Klasse. Nachrichten},
395--407.

\medskip
\noindent Howard, D. and Stachel, J. (eds.) 1989. {\it Einstein
and the History of General Relativity}, Boston, Basel, Berlin:
Birkh\"auser [Einstein Studies 1].

\medskip
\noindent Janssen, M.~1999. `Rotation as the Nemesis of Einstein's
{\it Entwurf} Theory'. in: \cite[Goenner et al. 1999,
127--157]{ES7}.

\medskip
\noindent Lecat, M. 1924. {\it Bibliographie de la Relativit\'e},
Bruxelles: Lamertin.

\medskip
\noindent Minkowski, Hermann. 1908. `Die Grundgleichungen f\"ur
die elektromagnetischen Vorg\"ange in bewegten K\"orpern'. {\it
K\"onigliche Gesellschaft der Wissenschaften zu G\"ottingen.
Mathematisch-physikalische Klasse. Nachrichten,} 53--111.

\medskip
\noindent Minkowski, H. 1909. `Raum und Zeit'. {\it Physikalische
Zeitschrift, 10}, 104--111.

\medskip
\noindent Norton, J. 1984. `How Einstein Found His Field
Equations'. {\it Historical Studies in the Physical Sciences}
14(1984), 253--316 [reprinted in \cite[Howard and Stachel 1989,
101--159]{ES1}].

\medskip
\noindent Pais, A. 1982. {\it `Subtle is the Lord...' The Science
and the Life of Albert Einstein}, Oxford: Oxford University Press.

\medskip
\noindent Ricci, G., and Levi-Civita, T.~1901. `M\'ethodes de
calcul diff\'erentiel absolu et leurs applications'. {\it
Mathematische Annalen, 54}, 125--201.

\medskip
\noindent Riemann, B. 1892. {\it Gesammelte mathematische Werke
und wissenschaftlicher Nachlass.} 2nd ed., Weber, H. (ed.)
Leipzig: Teubner.

\medskip
\noindent Renn, J., Sauer, T., Janssen, M., Norton, J. and
Stachel, J. forthcoming. {\it The Genesis of General Relativity}
(2~vols.), Dordrecht, Boston, London: Kluwer.

\medskip
\noindent Renn, J. and Sauer, T.~1999. `Heuristics and
Mathematical Representation in Einstein's Search for a
Gravitational Field Equation.' in: \cite[Goenner et al. 1999,
87--125]{ES7}.

\medskip
\noindent Sauer, T.~1999. `The Relativity of Discovery. Hilbert's
First Note on the Foundations of Physics'. {\it Archive for
History of Exact Sciences, 53}, 529--575.

\medskip
\noindent Scholz, E. (ed.)~2001. {\it Hermann Weyl's
Raum--Zeit--Materie and a General Introduction to His Scientific
Work.} Basel, Boston, Berlin: Birkh\"auser.

\medskip
\noindent Stachel, J. 1995. `History of Relativity', in: Brown,
L.M., Pais, A., and Pippard, B. (eds.) {\it Twentieth Century
Physics},  Philadelphia: Institute of Physics, Vol. I, 249--356.

\medskip
\noindent --- 2002. {\it Einstein from `B' to `Z'}. Boston, Basel,
Berlin: Birkh\"auser.

\medskip
\noindent Weyl, H.~1918. {\it Raum$\cdot$Zeit$\cdot$Materie.
Vorlesungen \"uber allgemeine Relativit\"atstheorie.} Berlin:
Springer [essential revisions in each of the later editions:
1919$^3$, 1921$^4$, 1923$^5$].

\end{document}